\documentclass[aps,pra,showpacs,superscriptaddress,twocolumn]{revtex4}
\usepackage{graphicx}
\usepackage{bm}
\usepackage{amsmath}
\usepackage{bbold}
\usepackage{hyperref}
\hypersetup{
     colorlinks=true,      
    linkcolor=blue,        
    citecolor=blue,        
    filecolor=magenta, 
    urlcolor=black          
}

\providecommand{\moy}[1]{\langle #1 \rangle}
\providecommand{\bra}[1]{\langle #1 \rvert}
\providecommand{\ket}[1]{\lvert #1 \rangle}
\providecommand{\braket}[2]{\langle #1 \rvert #2 \rangle}

\providecommand{\be}{\begin{equation}}
\providecommand{\ee}{\end{equation}}
\providecommand{\ba}{\begin{eqnarray}}
\providecommand{\ea}{\end{eqnarray}}

\begin{document}

\title{Testing the Clauser-Horne-Shimony-Holt inequality using observables with arbitrary spectrum}

\author{A. Ketterer}\email{andreas.ketterer@univ-paris-diderot.fr}
\affiliation{Laboratoire Mat\'eriaux et Ph\'enom\`enes Quantiques, Sorbonne Paris Cit\'e, Universit\'e Paris Diderot, CNRS UMR 7162, 75013 Paris, France}
\author{A. Keller}
\affiliation{Institut de Sciences Mol\'eculaires d'Orsay (CNRS), Universit\'e Paris-Sud 11, B\^{a}timent 350--Campus d'Orsay, 91405 Orsay Cedex, France}
\author{T. Coudreau}
\author{P. Milman}\email{perola.milman@univ-paris-diderot.fr}
\affiliation{Laboratoire Mat\'eriaux et Ph\'enom\`enes Quantiques, Sorbonne Paris Cit\'e, Universit\'e Paris Diderot, CNRS UMR 7162, 75013 Paris, France}
\date{\today}
 
\begin{abstract}
The Clauser-Horne-Shimony and Holt inequality applies when measurements with binary outcomes are performed on physical systems under the assumption of local realism.
Testing such inequalities in the quantum realm usually involves either measurements of two--valued quantum observables or  pre-defining a context dependent  binning procedure.
Here we establish the conditions to test the Clauser-Horne-Shimony and Holt inequality using any quantum observable.
Our result applies to observables with an arbitrary spectrum and no prior knowledge of their underlying Hilbert space's dimension is required.
Finally, we demonstrate the proposed general measurement strategy, that can be seen as positive operator valued measurements performed on the system, using the formalism of modular variables applied to the transverse degrees of freedom of single photons.
\end{abstract}
\vskip2pc 
 
\maketitle

\section{Introduction}

The Bell inequality was derived in 1964 \cite{BELL} in response to the A. Einstein, B. Podolski and N. Rosen  paradox \cite{EPR},  attempting to answer to the question: can quantum-mechanical description of physical reality be considered complete?  
By using the hypotheses of locality and realism, as used in \cite{EPR}, J. S. Bell, and later on,  J. Clauser, M. Horne, A. Shimony and R. Holt (CHSH) \cite{CHSH}  could establish an inequality that must be observed for measurements with binary outcomes  performed on classical systems, even when one admits the hypotheses of local hidden variables (LHV) evoked in \cite{EPR}. 
The interest of the CHSH inequality compared to the original Bell inequality is that it is easier to be tested experimentally, as demonstrated using quantum  two--dimensional systems, such as  the polarization of a photon \cite{ExpBell} or spin $1/2$ systems \cite{Ion}. 
The violation of the CHSH inequality can also serve as a benchmark for the quality of the entanglement produced by two--dimensional systems \cite{Horodecki, Weihs}.

The CHSH inequality involves correlations between measurements by two observers, $a$ and $b$ that can be expressed in terms of probabilities of obtaining two outcomes $\pm 1$. It was shown in \cite{Barut,Aspect} that this two valued binning, naturally present when observables with only two possible eigenvalues are measured, is essential in the  derivation of the CHSH inequality. Measuring and manipulating multi-valued and continuous quantities can also lead to inequalities testing the possibility of LHV  theories to explain non-locality \cite{Cavalcanti, Collins, Arnault}, but these are not the scope of the present work.  

In the present paper, we  address the following problem: what are the conditions an observable must satisfy to be used in a test of a CHSH inequality, which can be violated for some particular quantum states showing that quantum correlations are stronger than those arising form LHV theories?

This paper is organized as follows: in section \ref{sec1} we recall the CHSH inequality and then move to the main result of this paper, showing that {\it  any} observable  can be used to test a CHSH inequality. 
Our results can be interpreted using Positive Operator Valued Measurements (POVM) for each observer, naturally creating a binary statistics that can be used to  test, and for some states, violate, a CHSH inequality. We then discuss in section \ref{sec2} how our results can be used to test the CHSH inequality with no prior knowledge of the dimensionality of the Hilbert space of the measured observable. Section \ref{sec3a} is devoted to a demonstration of the proposed measurement strategy using the formalism of modular variables applied to transverse degrees of freedom of single photons. Finally, we conclude in Sec. \ref{sec:Conclusion}.

\section{The CHSH Inequality}\label{sec1}

In the CHSH inequality, two physical quantities, $A_{s_a}$ and $B_{s_b}$, are measured by observers $a$ and $b$ in different experimental settings $s_{\alpha}=\phi_{\alpha}^{(i)}$ where  $i$ determines a setting and $\alpha=a,b$.
To create a different experimental setting,  some  parameter of the experiment is modified, without altering the fact that only two values, $\pm 1$, can be assigned to the measured quantity.
The correlations appearing in the CHSH inequality are in  the form: 
\begin{equation}\label{correl}
 \langle A_{s_a}B_{s_b}\rangle=E(\phi_a^{(i)}, \phi_b^{(j)})=P_{++}^{i,j}+P_{--}^{i,j}-P_{+-}^{i,j}-P_{-+}^{i,j},
 \end{equation}
 where  $P_{k,l}^{i,j}$ is the joint conditional probability that $a$ and $b$ observe the value $k,l=\pm 1$ when $A_{\phi_{a}^{(i)}}$ and $B_{\phi_{b}^{(j)}}$ are measured in the $i,j$-th setting.
These joint probabilities fulfill the normalization condition $P_{++}^{i,j}+P_{--}^{i,j}+P_{+-}^{i,j}+P_{-+}^{i,j}=1$, for all settings $i$,$j$.

Using (\ref{correl}), the CHSH inequality can be written as:
\begin{widetext}
\begin{equation}\label{chsh}
|E(\phi_a^{(1)}, \phi_b^{(1)})+E(\phi_a^{(1)}, \phi_b^{(2)})+E(\phi_a^{(2)}, \phi_b^{(1)})-E(\phi_a^{(2)}, \phi_b^{(2)})| \rangle |\leq 2.
\end{equation}
\end{widetext}
It was shown in \cite{Barut, Aspect}  that the  CHSH inequality demands that the statistics of the measurement outcomes of  $A_{s_a}$ and $B_{s_b}$ are associated to some two--valued binning.
By such, we mean that it is mandatory that correlation functions can be written as in (\ref{correl}), which is a direct consequence of the fact that, for each party,  the expectation value of, say, $A_{s_a}$ can be written as  $\langle A_{s_a}\rangle=P_+-P_-$,  where $0 \leq P_{\pm} \leq 1$ and $P_++P_-=1$.
  
In the quantum formalism, the physical quantities  $A_{s_a}$ and $B_{s_b}$ become hermitian operators (observables): $A_{s_a} \rightarrow \hat A_{s_a}$ and $B_{s_b} \rightarrow \hat B_{s_b}$. The first observables that appeared as natural candidates  to test Eq. (\ref{chsh}) were quantum two-dimensional  ones, that can be constructed from the three Pauli matrices, $\hat \sigma_{\alpha}$, $\alpha=x,y,z$.
Measurement of the Pauli operators in any direction of space can only return the values $\pm 1$  with some probability that can be  directly associated to the probability $P_{\pm}$ of obtaining one of the two eigenstates of  $\hat A_{s_a}$ (or $\hat B_{s_b}$).

{Less intuitive examples of two--valued observables that can be measured for states in Hilbert spaces of higher dimensions can be constructed by dichotomization \cite{BANA, Reid1, Dicho,Jaksch1,Jaksch2}. This strategy allows the testing and violation of  the CHSH inequality  using a two-valued binning of experimentally accessible continuous variable measurements. One example of a dichotomic observable is the displaced parity operator whose expectation value can be directly associated with the Wigner function \cite{BANA, pseudo, teich,Kuzmich}. Furthermore, a test of the CHSH inequality for high dimensional discrete and continuous systems was proposed in \cite{Jaksch1} by binning local measurement outcomes. The same authors generalized the parity-based approach in \cite{BANA} in order to test the CHSH inequality in terms of a parametrized set of quasiprobability functions including the Wigner function as limiting case. Another state-dependent binning procedure was proposed in \cite{WENGER} by assigning binary probabilities to continuous measurement outcomes lying in some pre-defined regions of the phase space.
Finally, in \cite{Eberly}, a measurement strategy based on  auxiliary two-level systems was devised to transfer information of continuous variables states to observables with a two--valued spectrum.  }

The above strategies to  test  the CHSH inequality aim at explicitly identifying  $P_{\pm}$ to probabilities of  projective measurement outcomes. However, this leads  to excessively restrictive conditions to observables  $\hat A_{s_a}$ or $\hat B_{s_b}$, that are required to have specific spectral properties. Such restrictions not only reduce the space of states that can violate a CHSH inequality but also the possible experimental set-ups that are suitable to test it.

From the experimental point of view the ideal situation would be as follows: suppose  that  observers, $a$ and $b$ are capable of measuring  observables $\hat \chi_a$ and $\hat \chi_b$ with an arbitrary spectrum that is not necessarily known; is it possible to design a strategy for  $a$ and $b$ to test a CHSH inequality using measurements of such observables, without resorting to dichotomization or to a binning procedure? 
We show  that it is indeed possible to test a CHSH inequality using observables with an arbitrary spectrum if one realizes that  equation (\ref{chsh}) merely involves binary correlations between measurement results, and only the expectation values of observables are relevant, not their spectrum. 
This possibility is interesting not only because it renders CHSH inequality tests accessible to any experimental set-up, but also because it helps designing loophole-free experimental tests the CHSH inequality. 

In order to obtain a binary statistics in the quantum realm, requiring that observables $\hat A_{s_a}$ and $\hat B_{s_b}$ have a two-valued spectrum is a too strong demand, since  (\ref{chsh}) involves probabilities only. Testing (\ref{chsh}) demands designing experiments simply reproducing the same type of statistics as in (\ref{correl}) but such experiments can, in principle, involve systems with an arbitrary Hilbert space dimension. Thus, observables $\hat A_{s_a}$ and $\hat B_{s_b}$ must merely be such that 
\begin{equation}\label{qcorrel}
\left\langle \hat A_{\phi_a^{(i)}} \hat B_{ \phi_b^{(j)}}\right\rangle=E(\phi_a^{(i)}, \phi_b^{(j)})=P_{++}^{i,j}+P_{--}^{i,j}-P_{+-}^{i,j}-P_{-+}^{i,j},
\end{equation}
$\forall \, i,j$.
A combination of four such values with different settings $i,j$ ($i,j=1,2$), as in (\ref{chsh}), leads to a violation of the inequality (\ref{chsh}) for some entangled quantum states $\ket\Psi$.
We will see that the condition such observables must satisfy does not necessarily require a known {\it a priori} binning process, and that the previous cited examples of two--valued observables are particular cases of the general operators we obtain.

We start by  considering a single party operator, $\hat A_{s_a}$ and the same reasoning will be applied to  $\hat B_{s_b}$.
Since our basic requirement is  that $\langle\hat A_{s_a}   \rangle=P_+-P_-$ with $P_++P_-=1$, there exists a $\varphi$ such that $\langle\hat A_{s_a}   \rangle=\cos{\varphi}$,  $P_-=\sin^2{\frac{\varphi}{2}}$ and $P_+=\cos^2{\frac{\varphi}{2}}$.
Meaning that, if we are able to measure an arbitrary observable $\hat \chi_a$, with eigenstates defined as $\ket{\chi_a}$ and eigenvalues $\chi_a$ such that $\chi_a \  {\rm mod}\ 2\pi=\varphi$, we can construct any unitary operator $\hat D_{s_a}=\int_0^{2\pi}d\chi_a e^{i f(\chi_a)} \hat P(\chi_a)$, where  $f(\chi_a)$ is an arbitrary function and $\hat P(\chi_a)$ is the projector onto the eigenspace associated to a value of $e^{i f(\chi_a)}$. In the present paper, for simplifying reasons, we will focus on the case where $f(\chi_a)=\chi_a$. We thus define the real part of $\hat D_{s_a}$ as:
\begin{equation}\label{unitary}
\hat A_{s_a} =\int_0^{2\pi}d\chi_a \cos{\chi_a} \hat P(\chi_a).
\end{equation}

In (\ref{unitary})  we considered the most general case, where $\hat A_{s_a}$ has a continuous spectrum.
Operators with a discrete spectrum can be written in a form analogous to (\ref{unitary}) by replacing the integral by a sum over the discrete eigenstates of $\hat A_{s_a}$. For instance, Pauli matrices are a particular case of (\ref{unitary}).
In this case,  there are only two eigenstates $\ket{\chi_a^{(j)}}$, $j=0,1$,  with eigenvalues $\cos{(\theta+(-1)^{j}\pi/2)}$.
For any value of $\theta$, a proper normalization recovers the Pauli matrices with  spectrum $\pm 1$.

Our results can be interpreted in terms of a two valued POVM, $\hat E_{\pm}$, such that
\begin{align}
\hat E_{\pm} = \frac{1}{2}(\mathbb{1} \pm \hat A_{s_a}), 
\label{probabilities+-}
\end{align}
Indeed, since the spectrum of $\hat A_{s_a}$ is in
$[-1,1]$, $\hat E_{\pm}$ are positive operators, and by construction, 
$\hat E_- + \hat E_+ = \mathbb{1}$ and $\hat E_+ - \hat E_- = \hat A_{s_a}$. Therefore the probabilities 
 $P_{\pm} = \moy{\hat E_{\pm}}$ fulfill the required relations :  $P_+ - P_- = \moy{\hat A_{s_a}}$ and $P_- + P_+ = 1$. Using Eq. (\ref{unitary}),  we have that  $\hat A_{s_a}=(D_{s_a} + D_{s_a}^{\dagger})/2$ and the POVM  $\hat E_{\pm}$  can thus be written as:
\begin{align}
\hat E_{\pm}&=\frac{1}{4} (\mathbb 1\pm\hat D_{s_a}^{\dagger})(\mathbb 1\pm\hat D_{s_a}),
\label{eqn:POVM}
\end{align}
while the case of binary projective measurements is recovered when $\hat D_{s_a}$ is hermitian. 
Correlations between the outcomes of measurements of  $\hat A_{s_a}$ and  $\hat B_{s_b}$ in the form (\ref{unitary}) can be constructed as:
\begin{equation}\label{P++}
P_{k,l}=\frac{1}{4}\left (1+k \moy{\hat A_{s_a} }+l  \moy{\hat B_{s_b} }+{kl} \moy{\hat A_{s_a}\hat B_{s_b} }\right )\!\!,
\end{equation}
$k,l=\pm1$, directly leading to (\ref{qcorrel}), without the need of a prior choice of a  binning procedure, dramatically simplifying experimental tests of the CHSH inequality. This formulation is a special case of the more general fact that the expectation value of any bounded observable can be expressed in terms of a binary POVM \cite{HorodeckiPOVM}.
 
Considering such general form is interesting because it opens the perspective of testing CHSH inequalities in any measurable quantum system: once one knows how to measure the probability of finding a quantum system in the eigenstates $\ket{\chi_a}$  of an arbitrary measurable observable $\hat \chi_a$, operators as (\ref{unitary}) can be constructed, leading to the test of a CHSH inequality using (\ref{P++}). In many experimental systems, it appears that projective measurements  are not accessible. In \cite{Milman} it was suggested that parity measurements \cite{BANA} could lead to violation of the CHSH inequality. However, the state of the art of this set-up is far from enabling such projective measurements \cite{Wigner}, and only POVMs are accessible. 

A consequence of the presented results is that knowing the general form (\ref{unitary}) allows testing a CHSH inequality with no {\it a priori} information about the Hilbert space dimension of the system to be measured and on possible binning or dichotomization procedures.
By measuring an observable in the form (\ref{unitary}) in some experimentally accessible basis  $\ket{\chi_a}$, one is sure to be able to test a CHSH inequality without requiring any other information about the system. This point will be further discussed in the next section, where we consider a situation where one does not know, {\it a priori}, the Hilbert space dimension of the measured observable. 

\section{Illustration using measurements performed to observables of unknown dimension}\label{sec2}

In order to illustrate  this fact and the power of our results,  we consider the following situation:  suppose that each one of the observers $a$ and $b$ has a measuring apparatus that correlates some physical property of a quantum system to its position and then measures it.
The measurement of  the referred quantum property is thus indirectly  performed by collecting statistics about the quantum system's position.
This scenario is probably the most current in physics, corresponding to Stern-Gerlach (SG) type experiments \cite{SG}, measurements of the spectrum of a multi-mode field using a diffraction grating, the Mach-Zehnder interferometer, among many others.
If the referred physical property is associated to an observable with unknown  dimension, as was the case of the first SG experiment, no assumption can be made, {\it a priori}, on the possible outcomes of the position measurements.
In order to be even more general, one can suppose that the way correlations are created between position and the physical quantity to be measured is also unknown.
Thus, observers $a$ and $b$, when inspecting the position probability distribution that is measured by their apparatuses, have simply no idea of what type of results they can find.
Using the present formulation, even in this extremal case,  there exists a strategy enabling, in principle,  the violation of the CHSH inequality (\ref{chsh}).

This scenario can be expressed mathematically as follows: we consider first the case of a single party system and then extend our results to the bipartite case.
The initial state of the system is given by
\begin{equation}\label{initstate}
 \ket{\psi}\ket{x_o}=\int_{-\infty}^{\infty} ds f(s) \ket{s}\ket{x_o},
\end{equation}
where  $\ket{s}$ is a complete basis describing the property that is being measured.
$x_o$ is the position measurement result when no correlation is created  between position and some property of $\ket{\psi}$.
In the case of a SG experiment, for instance, $\ket{s}$ would be the different projections of the spin into a given axis.
Before the realization of this experiment, it was believed that this quantity could assume a continuity of values, and the description in (\ref{initstate}) is the most general one in this case.
Of course, after the SG experiment, a specific discrete form was inferred for $f(s)$.
Starting from (\ref{initstate}) we create  correlation between position and some property of  $\ket{\psi}$, that will be labeled $\psi_s$.
The system's  state becomes:
\begin{equation}\label{initstate2}
 \int_{-\infty}^{\infty} d\psi_s \ket{\psi_{s}}\ket{x(\psi_s)},
\end{equation}
where $\ket{\psi_s}=\int_{-\infty}^{\infty}ds f_{\psi_s}(s)\ket{s}$ and $\int_{-\infty}^{\infty}d\psi_s \ket{\psi_s}=\int_{-\infty}^{\infty} ds f(s) \ket{s}$.
In order to illustrate (\ref{initstate2}) we can think of the example of perfect correlation between position $\ket{s}$.
In this case, $f_{\psi_s}(s)=f(\psi_s)\delta(s-\psi_s)$.
This situation describes the SG experiment or diffraction by a grating.
Another situation would be the one of parity measurements, i.e., when position becomes correlated to the parity of the state.
Then, by writing $f(s)=f_o(s)+f_e(s)$, where $f_{o,e}(s)$ are the odd/even parts of $f(s)$,  the system's state is split in two according to its parity, and there are only two states $\ket{\psi_s}$: $f_{\psi_s}(s)=f_{o,e}(s)$.
Correlations between parity and position can be created in optical systems, for instance,  by using an orbital angular momentum sorter \cite{OAMsorter}.
The previous cases are extreme examples, illustrating either perfect or binary correlation.
Both extremes, and all the intermediate cases are covered by the present formalism, even of one does not  know what type of correlations are created between position and some property of the system.
Since $a$ and $b$ are realizing position measurements, in order to be able to test (\ref{chsh}), they can use an operator as (\ref{unitary}) in the position basis, irrespectively of the type of information position measurements bring from the quantum system.

An example of an unitary operator in the position basis is  $\hat U=\int_{-\infty}^{\infty}e^{i x \frac{2\pi}{\ell}}\ket{x}\bra{x}$, where $\ell$ is some number with the dimension of a length.
$\hat U$ is infinitely degenerated, since its eigenvalues are $\ell$ periodic.
The integral over the whole space can be replaced by an integral over one period and an infinite sum if one defines the following notation: let $\bar x=x\  {\rm  mod} \  \ell$ and $\ket{\tilde x}\bra{\tilde x}=\sum_{n=0}^{\infty}\ket{\bar x+\ell n}\bra{\bar x+\ell n}$ \cite{AHARONOV}.
Then, $\hat U$ can be written as:
\begin{equation}\label{Upos}
\hat U=\int_0^{\ell}d\bar x e^{i\bar x \frac{2\pi}{\ell}}\ket{\tilde x}\bra{\tilde x}.
\end{equation}
By identifying $\hat U$ with $\hat D_{s_a}$, we have that
\begin{eqnarray}\label{A}
\hat A_{s_a}&=&\int_0^{\ell}d\bar x \cos{\left(\bar x 2\pi /\ell\right)}\ket{\tilde x}\bra{\tilde x} \nonumber \\
&=&\int_0^{\ell/2}d\bar x \cos{(\bar x 2\pi /\ell)}\hat \sigma_z({\tilde x}),
\end{eqnarray}
where $\hat \sigma_z({\tilde x})=\ket{\tilde x}\bra{\tilde x}-\ket{\tilde x+\ell/2}\bra{\tilde x+\ell/2}$ \cite{PVG}.
This is an example of an  observable in the form (\ref{unitary}) in the position basis that can reveal the binary statistics required for testing CHSH inequalities.
 We now compute its expectation value using the general state (\ref{initstate2}).
For this, we re-express  (\ref{initstate2}) using that $x(\psi_{s_a})=\bar x(\psi_{s_a})+2\pi n(\psi_{s_a})$, leading to
\begin{equation}\label{expA}
\langle \hat A_{s_a} \rangle=\int_0^{\ell}d\bar x(\psi_{s_a}) \cos(\bar x(\psi_{s_a})2\pi /\ell)p(\bar x(\psi_{s_a})).
\end{equation}
where $p(\bar x(\psi_{s_a}))=\sum_{n(\psi_{s_a})} \int_{-\infty}^{\infty}ds_a |f_{\psi_{s_a}}(s_a)|^2$ and the limit of the sum can in principle go to infinity.
Expressions analog to (\ref{expA}) can be created to  $\langle \hat B_{s_b} \rangle$ and $\langle \hat A_{s_a}\hat  B_{s_b}\rangle$.
However, in this last case, one would have to compute the coincidence position measurements of $a$ and $b$.
In this bipartite case, the initial state is 
\begin{equation}\label{initstate3}
 \ket{\Psi}\ket{x_o}\ket{y_o}=\int_{-\infty}^{\infty}\int_{-\infty}^{\infty} ds_a ds_b f(s_a, s_b) \ket{s_a}\ket{ s_b}\ket{x_o}\ket{y_o},
\end{equation}
where $\ket{y_o}$ is the position of the particle detected by observer $b$.
Following the same principles as before, correlations between position and the physical quantities measured by $a$ and $b$ are given by:
\begin{equation}\label{initstate4}
 \int_{-\infty}^{\infty}  \int_{-\infty}^{\infty}d\psi_{s_a}d\psi_{s_b} \ket{\psi_{s_a}}\ket{x(\psi_{s_a})}\ket{\psi_{s_b}}\ket{y(\psi_{s_b})},
\end{equation}
where $\ket{\psi_{s_a}}\ket{\psi_{s_b}}=\int_{-\infty}^{\infty}\int_{-\infty}^{\infty}f_{\psi_{s_a}, \psi_{s_b}}(s_a, s_b)\ket{s_a}\ket{ s_b}ds_a ds_b$ and $\int_{-\infty}^{\infty}\int_{-\infty}^{\infty}d\psi_{s_a}d\psi_{s_b} \ket{\psi_{s_a}}\ket{\psi_{s_b}}=\int_{-\infty}^{\infty}\int_{-\infty}^{\infty} ds_a ds_b f(s_a, s_b) \ket{s_a}\ket{ s_b}=\ket{\Psi}$.
We can thus write  
\begin{widetext}
\begin{equation}\label{correcos} 
\langle \hat A_{s_a}\hat  B_{s_b}\rangle=\int_0^{\ell} \int_0^{\ell} d\bar x d\bar y (\psi_{s_a})d\bar y(\psi_{s_b}) \cos(\bar x(\psi_{s_a})2\pi /\ell) \cos(\bar y(\psi_{s_b})2\pi /\ell)p(\bar x(\psi_{s_a}),\bar y(\psi_{s_b})), 
\end{equation}
\end{widetext}
where $p(\bar x(\psi_{s_a}),\bar y(\psi_{s_b}))=\sum_{n(\psi_{s_a}), n(\psi_{s_b})}\int_{-\infty}^{\infty}\int_{-\infty}^{\infty}ds_a ds_b |f_{\psi_{s_a}, \psi_{s_b}}(s_a, s_b)|^2$ \cite{note}. We can notice that the presented results are quite general and do not depend on the specific form of the function $f_{\psi_{s_a}, \psi_{s_b}}(s_a, s_b)$, describing how correlations between the observable to be measured and position are created.

Eq. (\ref{correcos}) corresponds to an arbitrary experimental setting $s_a,s_b$.
In (\ref{chsh}) one must be able to change settings and compare correlations between the measurement results in different settings.
This can be done by applying  an unitary operation to the measurement apparatus, which is formally equivalent to applying it  to the initial state $\ket{\Psi}$ while keeping the measurement apparatus unchanged.
Correlations between measurements in different settings can be combined, as in (\ref{chsh}), so as to test the CHSH inequality for a system whose Hilbert space dimension is unknown.
In this scenario, if one supposes that observers blindly compute the quantity (\ref{correcos}) for each setting without necessarily observing the statistics of position measurements in detail, it would be possible to violate the CSHS inequality without discovering that the spin projections are two-valued quantities, in the case of a SG experiment.

We now discuss a specific example, showing how measurements on the transverse degrees of freedom of photon pairs can be handled using our ideas so as to lead to the violation of the CHSH inequality. 

\section{Example: CHSH inequality violation using the transverse degrees of freedom of photons} \label{sec3a}
\subsection{Modular variables formulation}\label{sec:ModVarTheo}
We illustrate our recipe for testing CHSH inequalities in arbitrary dimensions using the formalism of modular variables \cite{AHARONOV,PVG}. Therefore, we split the position and momentum operator into a discrete and a modular part:
\begin{eqnarray}
\hat x&=&\hat{\bar x}+\hat N \ell, \nonumber \\
\hat p&=&\hat{\bar p}+\hat M h/\ell, 
\end{eqnarray}
where $\hat N$ ($\hat M$) has integer eigenvalues, and $\hat{\bar x}= \hat x\ \text{mod}\ \ell$ ($\hat{\bar p}=\hat p\ \text{mod}\ h/\ell$) represents the modular position (momentum) operator with eigenvalues $\bar x \in [0,\ell[$ ($\bar p \in [0,h/\ell[$). As in Sec. \ref{sec2}, $\ell$ is an arbitrary length scale that can be chosen later on according to some property of the physical system under consideration. Since the modular position and momentum operator commute \cite{AHARONOV}, we can define a common set of eigenstates $|\{\bar x, \bar p\}\rangle =\sqrt{\frac{\ell}{h}}\sum_{n=-\infty}^{+\infty}
e^{i\bar{p} n \ell/\hbar }  |\bar x+ n \ell\rangle
= \sqrt{\frac{1}{\ell}}
e^{-i\bar p \bar x/\hbar}\sum_{m=-\infty}^{+\infty}
e^{-i2\pi m \bar x/\ell}
|\bar p+m h/\ell\rangle$, referred to as modular eigenstates. In the following, we will express all operators and states in terms of the modular basis.

In order to test the CHSH inequality using the position measurements, as exemplified in Sec.~\ref{sec2}, we use the operator (\ref{A}) which, in the modular basis, reads:
\begin{eqnarray}\label{Amodvar}
\hat A_{0}&=&\int_0^{\ell/2}d\bar x \int_0^{h/\ell}d\bar p \cos{(\bar x 2\pi /\ell)}\hat \sigma_z({\bar x,\bar p}),
\end{eqnarray}
where $\hat\sigma_z({\bar x,\bar p})=\ket{\{\bar x, \bar p\}}\bra{\{\bar x, \bar p\}}-\ket{\{\bar x+\ell/2, \bar p\}}\bra{\{\bar x+\ell/2, \bar p\}}$. Furthermore, to specify different measurement settings $\phi_a$ we define the transformed operator $\hat A_{\phi_a}= \hat U^\dagger(\phi_a)\hat A_{s_a} \hat U(\phi_a)$, using the continuous set of unitaries $\hat U(\phi_a)=e^{i \hat p^2 \frac{\ell^2}{h^2}\phi_a}$. Specifically, by choosing $\phi_a=\pi/2$, we get:
\begin{eqnarray}\label{AmodvarPi}
\hat A_{\frac{\pi}{2}}=\int_0^{\ell/2}d\bar x \int_0^{h/\ell}d\bar p \cos{(\bar x 2\pi /\ell-\bar p \ell/(2\hbar))}\hat \sigma_y({\bar x,\bar p}), \nonumber \\
\end{eqnarray}
where $\hat\sigma_y({\bar x,\bar p})=i \ket{\{\bar x+\ell/2, \bar p\}}\bra{\{\bar x, \bar p\}}-i\ket{\{\bar x, \bar p\}}\bra{\{\bar x+\ell/2, \bar p\}}$. Now, according to Sec.~\ref{sec1}, we use the operators (\ref{Amodvar}) and (\ref{AmodvarPi}), to define the Bell operator:
\begin{eqnarray} \label{ModVarBellOp}
\hat{\mathcal B}=\hat A_0\otimes\hat B_0+\hat A_0\otimes\hat B_{\frac{\pi}{2}}+\hat A_{\frac{\pi}{2}}\otimes\hat B_0-\hat A_{\frac{\pi}{2}}\otimes\hat B_{\frac{\pi}{2}}.
\end{eqnarray}
where we used capital letters to distinguish between system $A$ and $B$. Candidates of states for which the expectation value the operator~(\ref{ModVarBellOp}) violates the local-realism threshold of $2$ can be found by diagonalizing the operator $\hat{\mathcal B}(\bar x_a, \bar p_a,\bar x_b, \bar p_b)=\hat\sigma_z({\bar x_a,\bar p_a})\otimes \hat\sigma_z({\bar x_b,\bar p_b})+\hat\sigma_z({\bar x_a,\bar p_a})\otimes \hat\sigma_y({\bar x_b,\bar p_b})+\hat\sigma_y({\bar x_a,\bar p_a})\otimes \hat\sigma_z({\bar x_b,\bar p_b})-\hat\sigma_y({\bar x_a,\bar p_b})\otimes \hat\sigma_y({\bar x_b,\bar p_b})$, in the basis $\{\ket{\{\bar x_a+i \ell/2, \bar p_a\}}\otimes \ket{\{\bar x_b+j\ell /2, \bar p_b\}} \}$, where $i,j=0,1$. This yields two nonzero eigenvalues $\pm 2\sqrt 2$  and the corresponding eigenvectors:
\begin{widetext}
\begin{align}\label{NonNormalizableStateVio1}
\ket{\psi_\pm(\bar x_a, \bar p_a;\bar x_b, \bar p_b)} &= \frac{1}{N_{\pm}} [\ket{\{\bar x_a, \bar p_a\}} \ket{\{\bar x_b, \bar p_b\}} + \ket{\{\bar x_a+\ell/2, \bar p_a\}} \ket{\{\bar x_a+\ell/2, \bar p_a\}} \nonumber \\
&\pm i(\sqrt{2}\mp1)\left(\ket{\{\bar x_a, \bar p_a\}} \ket{\{\bar x_a+\ell/2, \bar p_a\}} + \ket{\{\bar x_a+\ell/2, \bar p_a\}}\ket{\{\bar x_a, \bar p_a\}}\right)],
\end{align}
\end{widetext}
where $N_{\pm}=2(2\mp\sqrt{2})^{1/2}$. These nonnormalizable states violate the CHSH inequality maximally for the choice $\bar x_{a/b} =0,\ell/2$ and $\bar p_{a/b}=0$, yielding $\moy{\hat{\mathcal B}}=\pm2\sqrt 2$. Furthermore, we can construct physically meaningful states that still yield a large amount of violation by continuously superposing the states in Eq.~(\ref{NonNormalizableStateVio1}) with localized wave packets centered sufficiently near to the positions for which the maximum violation occurs. An example of such a state is:
\begin{align}\label{RealisticStateViolation}
\ket{\Psi}=&\iint_0^{\ell/2}d\bar x_ad\bar x_b  \iint_0^{\ell}d\bar p_ad\bar p_b \nonumber \\
&\times f_{a}(\bar x_{a},\bar p_{a}) f_{b}(\bar x_{b},\bar p_{b}) \ket{\psi_+(\bar x_a, \bar p_a;\bar x_b, \bar p_b)} 
\end{align}
where $f_{a/b}(\bar x_{a/b},\bar p_{a/b})$ are normalized wave packets with support contained in $[0,\ell/2[\times[0,h/\ell[$. We note, that the wave function $\bra{\{\bar x_a,\bar p_a\}}\braket{\{\bar x_b,\bar p_b\}}{\Psi}$ on the whole modular space $[0,\ell[^2\times[0,h/\ell[^2$, is not equal to  $f_{a}(\bar x_{a},\bar p_{a}) f_{b}(\bar x_{b},\bar p_{b})$, but according to Eq.~(\ref{RealisticStateViolation}) also involves the definition of  $\ket{\psi(\bar x_a, \bar p_a;\bar x_b, \bar p_b)} $. In the following, we use, as example, the wave function of spatially entangled photon pairs that have passed through diffraction gratings. As we will show in the Sec.~\ref{sec:ExpImplePhotons}, such a wave function can, in the modular representation, be represented by a product of two Gaussians together with periodic boundary conditions on the domain $[0,\ell/2[\times[0,h/\ell[$:
\begin{align}
&f_{a/b}(\bar x_{a/b},\bar p_{a/b})\propto e^{-(\bar x_{a/b} - a_{\bar x})^2/(2\sigma_{\bar x}^2)} e^{-(\bar p_{a/b} - a_{\bar p})^2/(2\sigma_{\bar p}^2)},
\label{eq:ModVarfab}
\end{align}
where $\sigma_{\bar x/\bar p}$ and $a_{\bar x/\bar p}$ denote the width and position of the wave function in the modular position and momentum. 

\begin{figure}[t!]
\includegraphics[width=0.475\textwidth]{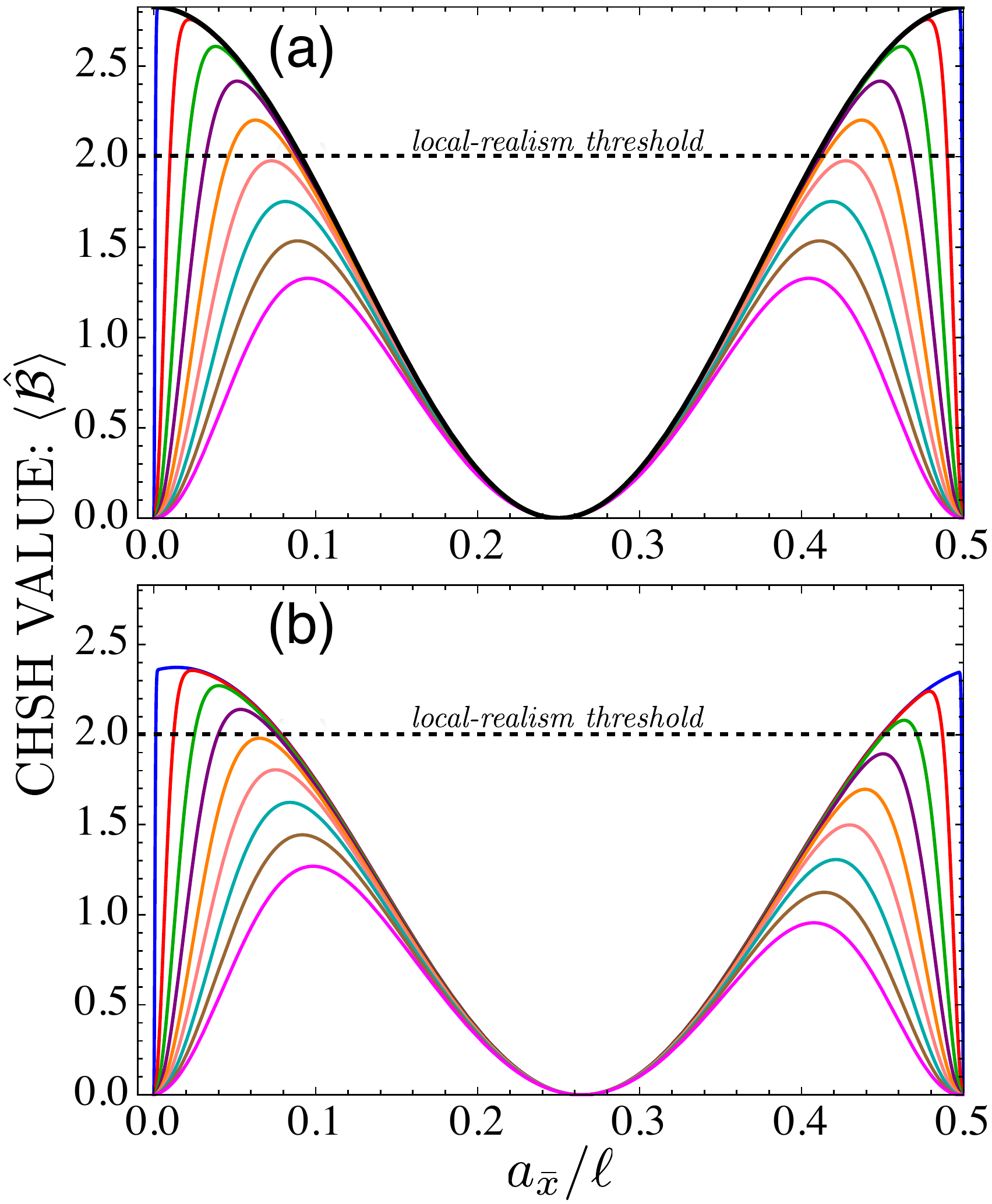}
\caption{(Color online) Expectation value of the Bell operator (\ref{ModVarBellOp}) according to the state (\ref{RealisticStateViolation}) as a function of $a_{\bar x}/\ell$, showing the violation of the CHSH inequality, for (a) a infinitely squeezed modular momentum distribution ($\sigma_{\bar p}\rightarrow0$) located at $a_{\bar p}=0$, and (b) for the values $\sigma_{\bar p}=0.1\ell/h$, $a_{\bar p}=0.1\ell/h$. Each line corresponds to a different width of the modular momentum distribution with  increasing order from the uppest to the lowest line: $\sigma_{\bar x}=0.001/\ell$ (blue, uppest), $\sigma_{\bar x}=0.01/\ell$ (red), $\sigma_{\bar x}=0.02/\ell$ (green), $\sigma_{\bar x}=0.03/\ell$ (purple), $\sigma_{\bar x}=0.04/\ell$ (orange), $\sigma_{\bar x}=0.05/\ell$ (pink), $\sigma_{\bar x}=0.06/\ell$ (cyan), $\sigma_{\bar x}=0.07/\ell$ (brown), $\sigma_{\bar x}=0.08/\ell$ (magenta,lowest). The black curve in plot (a) shows the function $2\sqrt 2 \cos^2{(2\pi a_{\bar x} /\ell)}$ and the black dashed line indicates the local-realism threshold.}
\label{fig:VioPlot}
\end{figure}

In Fig.~\ref{fig:VioPlot}, we present numerical results of $\moy{\hat{\mathcal B}}$ as a function of $a_{x}$, for different values of the parameters $\sigma_{x}$, $\sigma_{p}$ and $a_{p}$, showing the violation of the CHSH inequality. We see that, in the case of an infinitely squeezed modular momentum contribution ($\sigma_{p}\rightarrow 0$) centered at $a_{p}=0$ (see Fig. \ref{fig:VioPlot}(a)), the maximum of $\moy{\hat{\mathcal B}}$ passes the local-realism threshold if $\sigma_{\bar x}\lesssim0.049/\ell$, and approaches the value of maximal violation ($2\sqrt 2$) in the limit $\sigma_{\bar x}\rightarrow0$. Indeed, if we choose $f_{a/b}(\bar x_{a/b},\bar p_{a/b})\propto \delta{(\bar x-a_{\bar x})} \delta (\bar p)$ we get $\moy{\hat{\mathcal B}}=2\sqrt 2 \cos^2{(2\pi a_{\bar x} /\ell)}$, and thus $\moy{\hat{\mathcal B}}|_{a_{\bar x}=0}=2\sqrt 2$. Further on, for finite values of the modular momentum width $\sigma_{\bar p}$ and position $a_{\bar p}$ (see Fig. \ref{fig:VioPlot}(b)), the maximum of $\moy{\hat{\mathcal B}}$ is reduced but still violates the threshold $2\sqrt 2$ if $\sigma_{\bar x} \lesssim 0.039/\ell$. The asymmetry of $\moy{\hat{\mathcal B}}$ with respect to $a_{\bar x}$, in Fig.~\ref{fig:VioPlot}(b), originates from the $\bar p$ dependency of the \textit{cosine} in Eq.~(\ref{AmodvarPi}) which in the former case (see Fig. \ref{fig:VioPlot}(a)), for $a_{\bar p}=0$ and $\sigma_{\bar p}\rightarrow0$, does not play a role.

In the next Section, we will outline how the above described Bell test can be performed experimentally using measurements on the transverse degrees of freedom of photon pairs.

\subsection{Proposal of experimental implementation using the transverse degrees of freedom of photon pairs} \label{sec:ExpImplePhotons}
The mathematical structure of the spatial multimode field of a single photon reveals a perfect analogy with that of a single mode of the electromagnetic field containing a large number of photons \cite{TASCA}. In particular, due to the richness of optical elements and devices available in state of the art quantum optics experiments, the transverse degrees of freedom of a photon are an interesting playground for the implementation of quantum information applications \cite{TransverseDeutschJosza,ModVarWalborn}. In this section, we employ the transverse degrees of freedom of photon pairs in order to implement the above described test of the CHSH inequality (see Sec.~\ref{sec:ModVarTheo}).
\begin{figure}[h!]
\includegraphics[width=.475\textwidth]{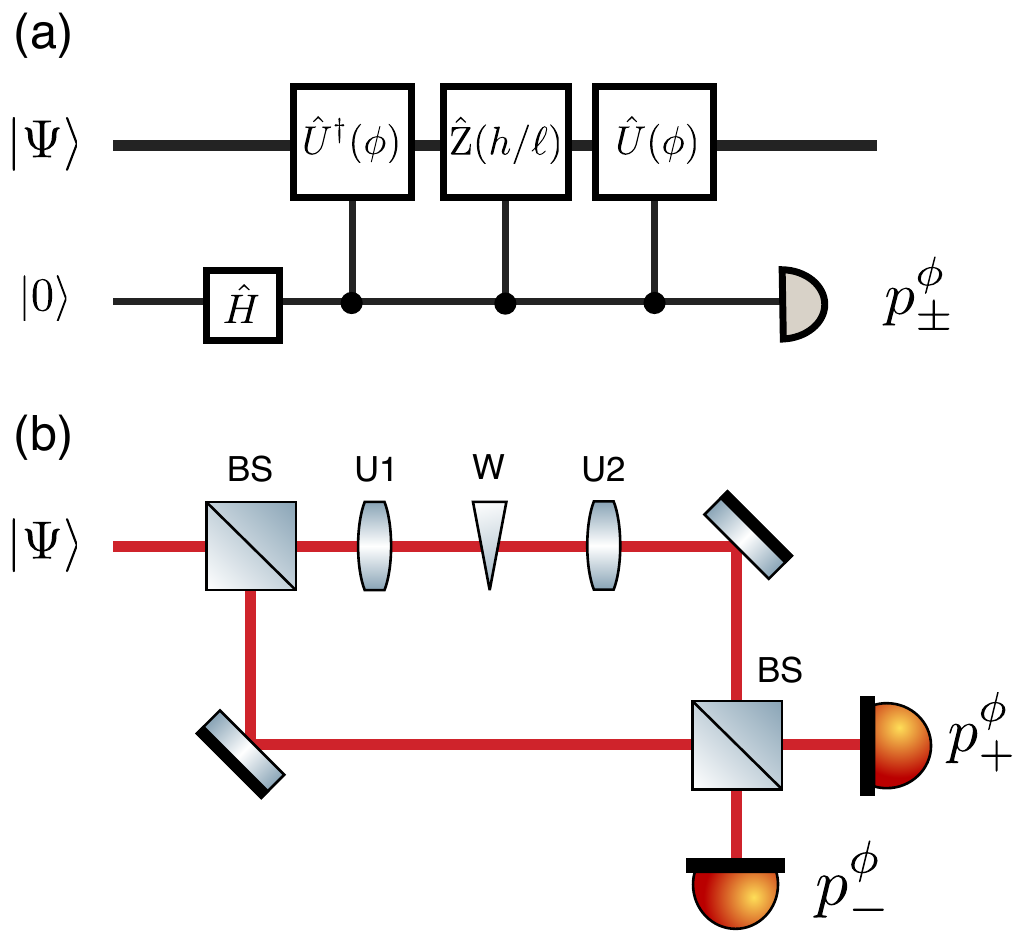}
\caption{(Color online) (a) Circuit diagram of an interferometer leading to the measurement of the expectation value of the operators $\hat A_\phi$, according to the state $\ket\Psi$ by coupling the continuous state $\ket\Psi$ to an ancilla qubit. $\hat H$ represents the Hadamard gate, and the operators $\hat{ \text Z}( \Delta p)=e^{i \hat x \Delta p/\hbar}$ and  $\hat{U}(\phi)=e^{i \hat p^2 \phi \ell^2/h^2}$ are implemented conditionally depending on the state of the ancilla. The probabilities $p^\phi_\pm$ are obtained by measuring the ancilla in the basis $\ket{\pm}=(\ket 0\pm\ket 1)/\sqrt 2$.  (b) Optical implementation of the above circuit using a balanced Mach-Zehnder type interferometer and linear optical elements acting on the transverse field of the photons. (U1) and (U2) indicate the free propagation of the photons and (W) a linear phase shift with $\Delta p=h/\ell$. The latter can be realized equivalently using a \textit{spatial light modulator} (SLM). Here, $p^\phi_\pm$ denote the photon counting probabilities for the detection of a photon in output $+$ or $-$.}
 \label{fig:MachZehnder} 
\end{figure} 

\subsubsection{Measurement of observables}
In the paraxial approximation it is convenient to describe the spatial degrees of freedom of a single photon state by a transverse wave function providing the probability amplitude for the detection of the photon in the transverse plane \cite{TASCA}. Moreover, it is possible to implement, on this transverse space, a universal set of unitary gates allowing for the approximate construction of all unitary operations by concatenating these gates. In order to construct the operators (\ref{Amodvar}) and (\ref{AmodvarPi}) we only need two unitary operators from the basic set of continuous unitary gates \cite{CVQuantumComputation}. Namely, the momentum shift operator $\hat{ \text Z}( \Delta p)=e^{i \hat x \Delta p/\hbar}$, and the single photon propagation $\hat{U}(\phi)=e^{i \hat p^2 \phi \ell^2/h^2}$, where $\hat x$ and $\hat p$ represent the transverse position and momentum operators of one of the two orthogonal transverse dimensions of the single photon, respectively. By combining $\hat U(\phi)$ and $\hat{\text Z}(\Delta p)$, we can define the operators:
\begin{align}
\hat A_0&=\hat{ \text Z}( h/\ell)+\hat{ \text Z}^\dagger( h/\ell),  \label{APhoton} \\
\hat A_{\frac{\pi}{2}}&=\hat{U}(\pi/2)\left[\hat{ \text Z}( h/\ell)+\hat{ \text Z}^\dagger( h/\ell) \right] \hat{U}^\dagger(\pi/2),
\label{ApiPhoton} 
\end{align}
where we set $\Delta p=h/\ell$ and $\phi=\pi/2$. Single unitaries, $\hat{ \text Z}( h/\ell)$ and $\hat{U}(\pi)$, can be implemented straightforwardly using linear optical elements and the free propagation of the photons \cite{TASCA}. However, for the sum of two unitaries one needs to employ an additional degree of freedom as the polarization of the photons, or their propagation direction, that can be, for instance, the two paths of an interferometer. In Fig.~\ref{fig:MachZehnder}, we demonstrate how the expectation value of the operators (\ref{APhoton}) and (\ref{ApiPhoton}) can be measured using a Mach-Zehnder interferometer setup, as discussed in previous works \cite{MachZehn1,MachZehn2}. We will see that this can be done by simply counting the arrival of photons in the outputs of the interferometer yielding the corresponding photon count probabilities:
\begin{align}
p^\phi_{\pm}=\frac{1}{2}\left(1\pm \text{Re}\moy{\hat{U}(\phi)\hat{ \text Z}( h/\ell) \hat{U}^\dagger(\phi)}\right),
\label{PhotonProbabilities} 
\end{align}
with $\phi=0,\pi$, and hence:
\begin{align}
\moy{\hat A_\phi}=p^\phi_+-p^\phi_-.
\label{ExpecAProbabilities} 
\end{align}
Note that the interferometers simply serve as strategy allowing the performance of the corresponding POVM's, $\hat E^\phi_{\pm}=\frac{1}{2}\left( \mathbb 1 \pm \text{Re}[ \hat{U}(\phi) \hat{ \text Z}( h/\ell) \hat{U}^\dagger(\phi)]\right)$ (compare with Eq.~(\ref{eqn:POVM})), on the transverse field of the photons.
Further on, the measurement of the products $\moy{\hat A_\phi \otimes \hat A_{\phi'}}$, as required in the CHSH inequality, can be preformed in an equivalent manner, using two analogous Mach-Zehnder interferometers, by counting the photon coincidences in the four outputs of the two interferometers, leading to the coincidence probabilities $P^{\phi,\phi'}_{k,l}=\moy{\hat E^{\phi,\phi'}_{k,l}}$, with the two-partite POVM (see also Eq.~(\ref{P++})):
\begin{align}
\hat E^{\phi,\phi'}_{k,l}&=\frac{1}{4}\left(\mathbb 1\otimes \mathbb 1+ l\ \hat A_{\phi}\otimes \mathbb 1+ k\ \mathbb 1\otimes \hat B_{\phi'} +kl\  \hat A_{\phi}\otimes \hat B_{\phi'} \right),
\label{PhotonProbabilities} 
\end{align}
for $k,l=\pm1$.
\begin{figure}[t!]
\includegraphics[width=0.49\textwidth]{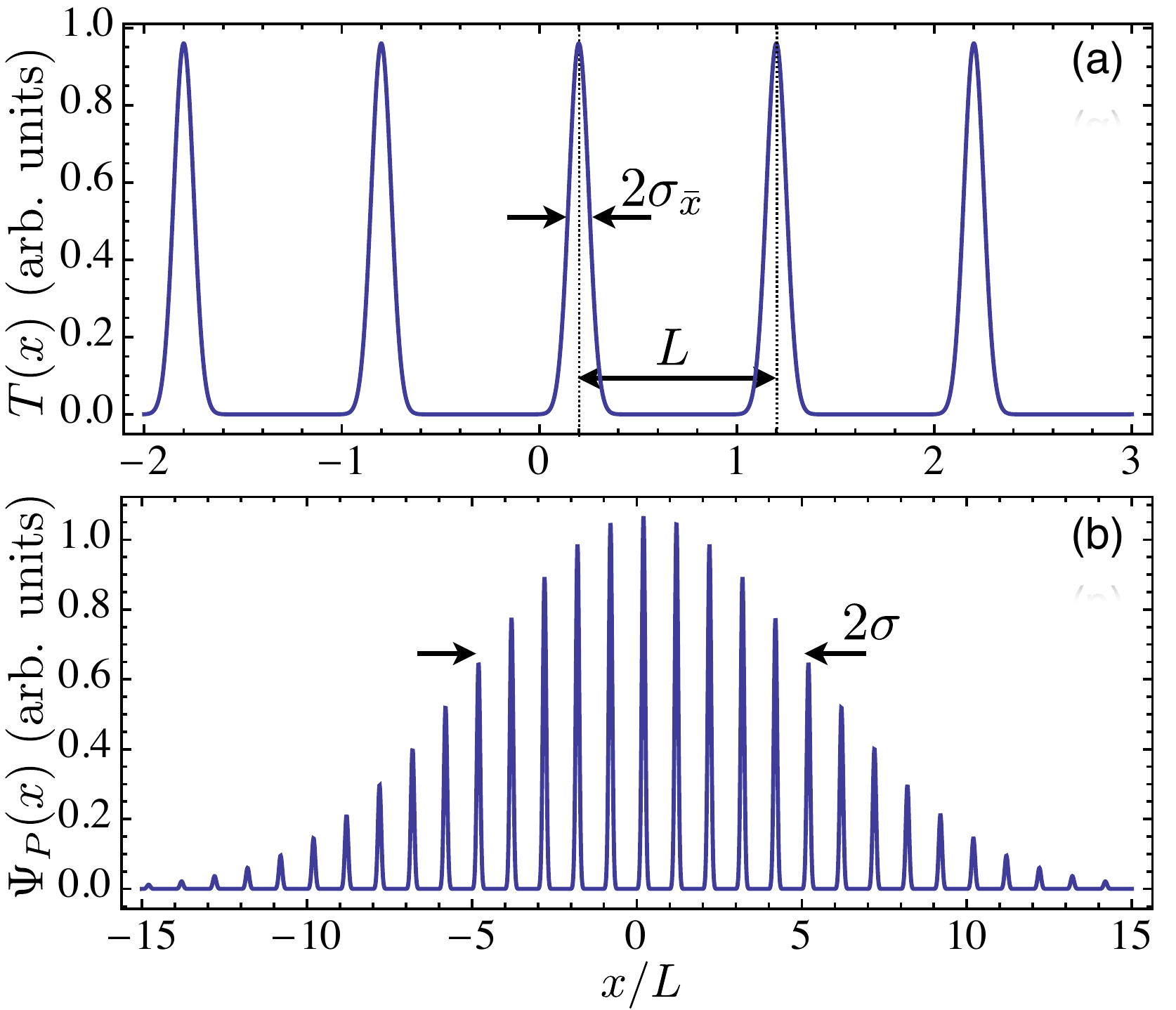}
\caption{(Color online) Examples of the plots of (a) the grating transmission function $T(x)$ with slit distance $L$ and $\kappa$ such that $\sigma_{\bar x}=0.01/L$, and (b) the single photon transverse wave function directly after it has passed through the diffraction grating defined by $T(x)$ and the Gaussian envelope with width $\sigma=5/L$. A plot of the $\bar x$-dependent part of the modular wave function $\widetilde\Psi_P(\bar x,\bar p)$ [see Eq.~(\ref{eq:GaussianCombModVar1})], namely $\widetilde T(\bar x)$, is given by that of $T(x)$ restricted to the domain $[0,L[$. }
 \label{fig:PlotInterference} 
\end{figure} 

\subsubsection{Preparing spatially entangled states}
Next, we explain how entangled states as (\ref{RealisticStateViolation}) can be produced experimentally using the spatial degrees of freedom of photon pairs. For that, we first show how to engineer single photon states in the form $\ket{f}=\int_0^{\ell/2}d\bar x \int_0^{h/\ell}d\bar p  f(\bar x,\bar p) \ket{\{\bar x,\bar p\}}$ or $\ket{\overline{f}}=\int_0^{\ell/2}d\bar x \int_0^{h/\ell}d\bar p  f(\bar x,\bar p) \ket{\{\bar x+\ell/2,\bar p\}}$, with $f(\bar x,\bar p)$ chosen according to Eq.~(\ref{eq:ModVarfab}).
 
Let us consider the transverse wave function of a single photon directly after it has passed through an infinitely extended diffraction grating with slit distance $L$ \cite{OpticsExpSinglePhotonInt}:  
 \begin{align}\label{eq:GaussianComb}
\Psi_{P}(x)&= T(x) f_G(x).
\end{align}
Here, $T(x)=\sum_{m=-\infty}^\infty c_m e^{i 2 \pi m x/L}$ is the $L$-periodic grating transmission function, defined through the Fourier coefficients $c_m$, such that $\sum_{m=-\infty}^\infty |c_m|^2=1$, and $f_G(x)=e^{-x^2/(2\sigma^2)}/(\sigma \pi)$ a Gaussian with width $\sigma$ accounting for the finite extension of the single photon wave packet. An example of $T(x)$ and $\Psi_{P}(x)$, for $c_m=e^{-m^2\kappa^2 /2}$, is shown in Fig. \ref{fig:PlotInterference}(a) and (b), respectively. We see that, for this choice of the coefficients $c_m$, $T(x)$ represents a comb of Gaussians with width $\sigma_{\bar x}=\kappa^2L^2/(2\pi)^2$, and $f_G(x)$ a Gaussian envelope thereof. 

Further on, if we assume $L/\sigma\ll1$ the approximation $e^{-(\bar x+n L)^2/(2 \sigma^2)}\approx e^{-(nL)^2/(2 \sigma^2)}$ becomes permissible and Eq.~(\ref{eq:GaussianComb}) transforms to the modular representation as:
\begin{eqnarray}\label{eq:GaussianCombModVar1}
\widetilde\Psi_{P}(\bar x,\bar p)=\sqrt{\frac{h}{\ell}} \ \widetilde T(\bar x) \widetilde C{(\bar p)} , 
\end{eqnarray}
where we set $\ell=2L$, $\widetilde T(\bar x)=\sum_{m=-\infty}^\infty c_m e^{i 4 \pi m \bar x/\ell}$ is defined by restricting $T(x)$ to the domain $[0,\ell/2[$ and $\widetilde C(\bar p)=\frac{1}{\sigma \pi}\sum_{n=-\infty}^\infty e^{-(n\ell)^2/ (2\sigma^2)} e^{i n \bar p \ell/\hbar}$ represents a comb of Gaussians in momentum space with width $\sigma_{\bar p}=h^2/(2\pi \sigma)^2$ restricted to the domain $[0,h/\ell[$. Equation~(\ref{eq:GaussianCombModVar1}) is thus nothing but the product of two Gaussians with width $\sigma_{\bar x}$ and $\sigma_{\bar p}$, respectively, and periodic boundaries on the domain $[0,\ell/2[\times[0,h/\ell[$ (compare with Eq.~(\ref{eq:ModVarfab})). Hence, by tuning the parameters $\kappa$ and $\sigma$, we are able to generate the states $\ket{f}$ or $\ket{\overline{f}}$. Thereby, the theoretical limit $\sigma_{\bar p} \rightarrow 0$, as assumed in the calculations leading to the results presented in  Fig.~\ref{fig:VioPlot}(a), of an infinitely squeezed modular momentum (see Eq.~(\ref{eq:ModVarfab})), corresponds to the idealized situation of an infinitely extended single photon transverse wave function ($\sigma \rightarrow \infty$).

Experimentally, the ratio of $\kappa/\sigma$ can be modified by adjusting the slit width and distance of the diffraction grating that is used to create the interference pattern. The transverse location of the grating can be associated to the position of the interference peaks of the photon's wave function (\ref{eq:GaussianComb}) with respect to its envelope (see Fig.~\ref{fig:PlotInterference}(b)), and thus to the value of $a_{\bar x}$. And finally, a nonzero mean modular momentum $a_{\bar p}$, as assumed in Fig. \ref{fig:VioPlot}(b), can be achieved by imprinting a position dependent phase $e^{i a_{\bar p} n \ell/\hbar}$, with $n\ell=x-\bar x$, on the transverse field of the photon (\ref{eq:GaussianComb}). The latter is readily realized using a \textit{spatial light modulator} (SLM) whose action on the photon's transverse wave function can be described as $e^{if(\hat x)}$ with a user-defined function $f(x)$. \\

\begin{figure}[t]
\includegraphics[width=0.49\textwidth]{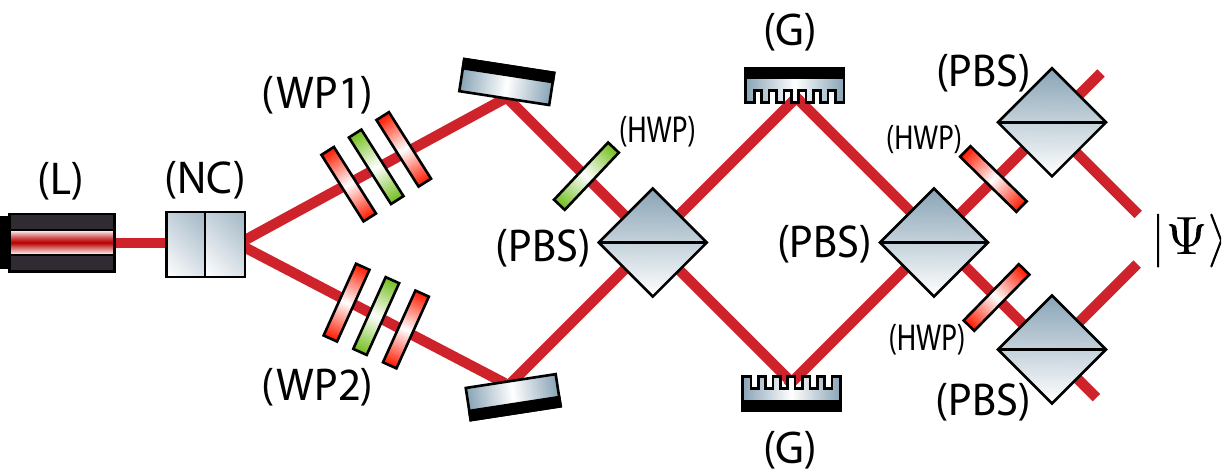}
\caption{(Color online) Proposed experimental setup to create a spatially entangled two photon state $\ket\Psi$ of the form (\ref{NonNormalizableStateVio1}). L: laser, NC: nonlinear crystals (type I), HWP: half-wave plate, WP1/WP2: combination of half- and quater-wave plates, PBS: polarizing beam splitter, G: diffraction grating.}
 \label{fig:StateGenerator} 
\end{figure}

Now, knowing how to prepare the single photon states $\ket{f}$ and $\ket{\bar f}$, we can use a technique that was previously proposed in \cite{Carolina}, in order to create the spatially entangled state (\ref{RealisticStateViolation}) of a photon pair. To this end, we first use spontaneous parametric downconversion to produce a polarization entangled state $\frac{1}{\sqrt 2}(\ket{\text H}_a\ket{\text H}_b+\ket{\text V}_a\ket{\text V}_b)$, by pumping two adjacent nonlinear crystals (type I) with a laser beam polarized at $45^\circ$, and, subsequently, bring it into the form:
\begin{align}\label{PlarizationEntangledState}
\ket{\psi_{pol}} =& \frac{1}{N_{\pm}} \Big[\ket{\text H}_a \ket{\text H}_b + \ket{\text V}_a \ket{\text V}_b  \nonumber \\
& \pm i(\sqrt{2}\mp1)\left(\ket{\text H}_a \ket{\text V}_b + \ket{\text V}_a \ket{\text H}_b\right)\Big],
\end{align}
where $N_{\pm}=2 \sqrt{2-\sqrt{2}}$, by the correct application of local unitary transformations on each of the entangled photons, realized by combinations of half- and quater-wave plates, as shown in Fig.~\ref{fig:StateGenerator}. Thereby, the transverse wave function of the two photons remains in the separable Gaussian state $\ket{\psi_g}_a\ket{\psi_g}_b$. Next, we have to swap the entanglement from the polarization to the transverse degrees of freedom of the photons according to $\ket{\text H}_{a/b}\ket{\psi_G}_{a/b}\rightarrow \ket{\text H}_{a/b} \ket f_{a/b}$ and $\ket{\text V}_{a/b}\ket{\psi_G}_{a/b}\rightarrow \ket{\text V}_{a/b}\ket{\bar f}_{a/b}$. This swapping is realized experimentally using an interferometer with polarizing beam splitters (see Fig. \ref{fig:StateGenerator}), such that, depending on the polarization of the photons, the operation $\ket{\psi_G}\rightarrow  \ket f$ or $\ket{\psi_G}\rightarrow \ket{\bar f}$, with the help of single photon diffraction gratings, can be performed. One of the two photons must be rotated before entering the interferometer, in order to assure that initially H-polarized $a$-photons will go through the same arm of the interferometer as H-polarized $b$-photons. Finally, one uses half-wave plates oriented at $\pi/8$ in the path of both photons and polarizing beam splitters in order to factorize the polarization from the transverse degree of freedom yielding the desired state (\ref{RealisticStateViolation}) conditionally with a $50\%$ probability.

\section{Discussion and Conclusion} \label{sec:Conclusion}

We now discuss the natural binning  procedure that is realized by the operator (\ref{unitary}). Consider the eigenstates with maximum (minimum)  eigenvalue of the observable $\hat A_{s_a}$, denoted as $\ket{\bar 1 (\bar 0)}$. In a realistic situation such states have a normalizable wave function with nonzero width $\epsilon$, which lead, in the limit $\epsilon \rightarrow 0$, to the expectation value $ \langle \hat  A_{s_a} \rangle=\bra{\bar 1 (\bar 0)}\hat A_{s_a} \ket{\bar 1 (\bar 0)} =\pm1$. 
When considering quantum systems of arbitrary dimension, it is clear that, due to degeneracies, an infinity of states, to which we refer to as $\{\ket{\alpha}\}$, can lead to the same value of $P_+-P_-$.
These states are associated to the same expectation value $\langle \hat A_{s_a}\rangle$, and can be regarded equivalent as far as this quantity is concerned.
 In this sense, they are also equivalent to a state $\ket{\bar \alpha}= \cos{\alpha}\ket{\bar 0}+e^{i\beta}\sin{\alpha}\ket{\bar 1}$, where $\beta$ is an arbitrary phase, such that 
\begin{equation}
\langle \hat A_{s_a}\rangle_{\{\ket{\alpha}\}}\!\!=\!\!P_+-P_-\!\!=\!\langle\hat A_{s_a}\rangle_{\ket{\bar \alpha}}\!\!=\cos^2{\alpha}-\sin^2{\alpha}\!=\!\cos{(2\alpha)}.
\end{equation}
Simple examples of such states are $\ket{\theta +2\pi n} $, where $n$ is an integer. These results can be straightforwardly extended to the correlations $\langle\hat A_{s_a}\hat B_{s_b}\rangle$.

In conclusion, we provided the general conditions observables with an arbitrary spectrum must satisfy to be suitable to test a CHSH type inequality.   
We have shown that dichotomic observables are a particular case of the formal solution developed here.
We also detailed the situation where observers have access to position measurements only and position is correlated in an unknown way to some physical property of the system to be measured.
 Our results help creating a general environment for CHSH inequalities tests without the need of complicated state dependent binning procedures or prior knowledge of the physical properties of the system, rendering CHSH inequalities tests accessible to a broader class of experimental systems and significantly simplifying their realization. An interesting perspective is applying the same principles developed here to other type of inequalities valid under the assumption of  local realism, as  \cite{Cavalcanti, Collins, Arnault, Garg}. 

\acknowledgments
The authors acknowledge financial support by ANR/CNPq HIDE and  ANR COMB.
The authors are indebted to P.~Grangier for inspiring discussions, and to P.H.S.~Ribeiro, S.P.~Walborn, J.-M. Raimond, M. Brune, J. Laurat and N. Treps for fruitful comments.


\begin{thebibliography}{99}
\bibitem{note} In practical situations, position measurements are realized in a given finite interval defined as $2l$, even  in the case where the spectra $s_a$ and $s_b$ are arguably infinite.
We can thus set $\Delta x>>2l$, which is equivalent to consider $n(\psi_{s_a}), n(\psi_{s_b})=0$  and $x(\psi_{s_a})=\bar x$ and  $y(\psi_{s_b})=\bar y$, that will simplify the notation and interpretation of the results.
\end{thebibliography}
\end{document}